\documentclass[a4paper,11pt]{article}
\pdfoutput=1 % if your are submitting a pdflatex (i.e. if you have
\usepackage{jcappub} % for details on the use of the package, please
\usepackage[T1]{fontenc} % if needed
\usepackage[export]{adjustbox}
\usepackage{subfigure,bbm}
\usepackage{graphicx}% Include figure files
\usepackage{dcolumn}% Align table columns on decimal point
\usepackage{bm,natbib}

\title{\boldmath On Reflectivity of Quantum Black Hole Horizons}

\author[a,b,c]{Naritaka Oshita}
\author[a,d,e]{Qingwen Wang}
\author[a,d,e]{Niayesh Afshordi}

\affiliation[a]{Perimeter Institute For Theoretical Physics, 31 Caroline St N, Waterloo, Canada}
\affiliation[b]{Research Center for the Early Universe (RESCEU), Graduate School
  of Science,\\ The University of Tokyo, Tokyo 113-0033, Japan}
\affiliation[c]{Department of Physics, Graduate School of Science,\\ The University of Tokyo, Tokyo 113-0033, Japan}
\affiliation[d]{Department of Physics and Astronomy, University of Waterloo, 200 University Ave W, N2L 3G1, Waterloo, Canada}
\affiliation[e]{Waterloo Centre for Astrophysics, University of Waterloo, Waterloo, ON, N2L 3G1, Canada}

\emailAdd{noshita@pitp.ca}
\emailAdd{qwang@pitp.ca}
\emailAdd{nafshordi@pitp.ca}

\abstract{We study the reflectivity of quantum black hole (BH) horizons using detailed balance and fluctuation-dissipation theorem, finding a universal flux reflectivity given by the Boltzmann factor ${\mathcal R} = \exp\left(-{\hbar |\tilde{\omega}| \over k T_{\rm H}}\right)$, where $\tilde{\omega}$ is frequency in the horizon frame and $T_{\rm H}$ is the Hawking temperature. This implies CP-symmetry (or $\mathbbm{RP}^3$ topology)  of the extended BH spacetime. We then briefly discuss related physical implications:  We predict echoes in the ringdown of Kerr BHs, but they do not exhibit ergoregion instability. The viscosity in the membrane paradigm is modified to $\eta = \frac{c^3}{16\pi G}\tanh\left({\hbar |\tilde{\omega}| \over 4 k T_{\rm H}}\right)$, only approaching General Relativistic value at high frequencies.}

\begin{document}
\maketitle
\flushbottom

\section{Introduction}
Black holes (BHs) are some of the most mysterious objects in the Universe. General Relativity predicts that they are surrounded by {\it event horizons} -- boundaries from inside which no signal can reach outside observers. Nevertheless, quantum effects are expected to lead to thermal radiation at Hawking temperature, and eventually lead to their evaporation \cite{Hawking:1974rv}. 
% Susskind \textit{et al.} \cite{Susskind:1993if} proposed the \textit{BH complementarity} in which the stretched horizon, located at about a Planck length outside the event horizon, can exhibit all the properties necessary to describe physics seen by the infalling observer. A related idea --membrane paradigm -- had been proposed earlier by Thorne \textit{et al.} \cite{Thorne:1986iy} and by Damour \cite{PhysRevD.18.3598}.

This evaporation process, however, has proved controversial over the past four decades, as the fate of information that falls into the BH remains illusive (e.g., \cite{Hawking:1976ra,Giddings:2006sj,Mathur:2008wi,Almheiri:2012rt}). In this context, a number of conjectures regarding the quantum aspects of BHs have been proposed which suggest quantum effects drastically change the near-horizon structure. For example, fuzzball horizonless geometries have been considered as microstates of black holes in string theory \cite{Mathur:2005zp}, while gravastars are suggested to emerge as a result of conformal anomaly (e.g., \cite{Mottola:2011ud,Kawai:2017txu}).  Almheiri \textit{et al.} \cite{Almheiri:2012rt} combined the holographic principle and locality with the monogamy of quantum entanglement to conclude that highly energetic quanta are excited near the horizon (aka {\it firewall}). Such near-horizon modifications are also proposed as ways to account for dark energy scale \cite{PrescodWeinstein:2009mp}, and BH entropy \cite{Saravani:2012is}. It is even suggested that such modifications can extend far from the horizon, if one insists on a ``non-violent unitarization'' framework \cite{Giddings:2017mym}. 
% Furthermore, it is conjectured that the area of BH horizon may be quantized in Planck area and the BH possesses discrete energy levels \cite{Bekenstein:1974jk,Bekenstein:1995ju}. 

Although several arguments have been proposed from the different viewpoints to grasp the essence of quantum BHs, the Bekenstein-Hawking entropy \cite{Bekenstein:1972tm,Bekenstein:1973ur,Gibbons:1976ue} may be universal for almost all of them since macroscopic thermodynamic quantities are independent of microscopic details.

In this paper, we discuss another possible universal quantity --surface reflectivity of quantum BHs-- and show that -- {\it {from independent aspects of}} detailed balance, the fluctuation-dissipation theorem, and CP-symmetry of the BH final state -- the reflectivity is given by the thermal Boltzmann factor. We then briefly discuss the physical implications for the late-time ringdown of gravitational waves from a spinning quantum BHs, as well as for ergoregion stability and viscosity in the membrane paradigm. A companion paper studies the prediction for ringdown echoes in more detail \cite{echo_QBH}. {The organization of the paper is as follows:
In the next section we discuss the derivation of Boltzmann reflectivity from three independent aspects. In addition, we revisit the membrane paradigm to investigate how our proposal changes the original viscosity predicted by the membrane paradigm. In Sec. \ref{sec:ECHO} we numerically investigate the GW echoes  emitted by a spinning BH in the Boltzmann reflectivity model. We also discuss how the Boltzmann reflectivity suppresses the ergoregion instability. The final section is devoted to the conclusion.}

\section{Boltzmann reflectivity}
\subsection{Detailed Balance}
% For the consistency between the second law of thermodynamics and a system including a BH, Bekenstein conjectured that \cite{Bekenstein:1972tm,Bekenstein:1973ur} a BH possesses its gravitational entropy, $S$, proportional to the area of event horizon, $S \propto A$, and Gibbons and Hawking gives the precise form of the entropy as $S = A/(4\ell_{\text{Pl}}^2)$ (Bekenstein-Hawking entropy) based on the (semiclassical) Euclideian path integral method \cite{Gibbons:1976ue}. The form of the Bekenstein-Hawking entropy implies that the event horizon of BH is quantized with the size of Planck area
% \begin{equation}
% A = N \times \alpha \ell_{\text{Pl}}^2,
% \end{equation}
% where $N$ is an integer and $\alpha$ is a constant of the order of unity. Since the area of event horizon is determined by the interior energy, $A=4 \pi r_g^2$ with $r_g \equiv 2 GM$, the quantization of horizon area means that the energy of BH is discrete and the $n$-th energy level $\Delta M_n$ is given by
% \begin{equation}
% \Delta M_n = n \times \frac{\alpha}{4} T_{\rm H}.
% \end{equation}
% If this conjecture is correct, BHs may be regarded as a quantum system with discrete energy levels.
From a quantum mechanical point of view, we can consider an isolated BH as an excited multilevel quantum system (e.g., a giant atom), which de-excites by emitting Hawking radiation. 
%{This picture is supported by for example the quantization of BH horizon area \cite{Bekenstein:1995ju}, in which a BH can be regarded as a multi-level quantum system with a energy gap of $\sim \hbar c^3/GM$.} 
We will now show that gravitational waves (GWs) infalling into a BH must be reflected near the horizon with the Boltzmann factor ${\mathcal R} = \exp\left(- \frac{\hbar \tilde{\omega}}{k T_{\rm H}}\right)$, where $\tilde{\omega}$ is the near-horizon frequency, and $T_{\rm H}$ is the Hawking temperature. 

Let us suppose that ingoing (large amplitude {and low-frequency}) GWs, of which the spectral energy density is denoted by $\rho (\tilde{\omega})$,  can stimulate the BH in a perturbative manner, and excite the multi-level quantum system from state j to state k, where $E_k-E_j = \hbar \tilde{\omega} >0$.
{Here we assume that the frequency of incoming gravitons is comparable to or lower than the typical energy scale of transition $\hbar \omega \lesssim k_{\rm B} T_{\rm H}$. In this situation, the lower-frequency gravitons just weakly coupled to the quantum BH, and according to the Fermi's golden rule, only the initial and final quantum states are involved in the quantum transition, e.g. stimulated emission. Therefore, in the following we discuss the transition between two energy levels by only taking into account the initial and final quantum states, but it does not mean that we regard the quantum BH as a two-level quantum system.}
The reflectivity of the BH can be characterized by the Einstein coefficients for spontaneous emission, stimulated emission, and absorption, denoted by $A_{kj} (\tilde{\omega})$, $B_{kj} (\tilde{\omega})$, and $B_{jk} (\tilde{\omega})$, respectively \cite{1917PhyZ...18..121E}. {The net rate of transition from state j to state k is given by:
\begin{equation}
R_{j \to k} - R_{k \to j}= B_{jk} n_j \rho -A_{kj} n_k -B_{kj} n_k \rho,
\label{detailed}
\end{equation}
where $R_{j \to k}$ ($R_{k \to j}$) is rate of transition from state j to k (k to j). The detailed balance condition  $R_{j \to k} = R_{k \to j}$ would guarantee that the BH remains in thermal equilibrium with a blackbody radiation $\rho(\tilde{\omega})$ given by: }
\begin{equation}
\rho_{\rm BB}(\tilde{\omega}) = \frac{2 \hbar\tilde{\omega}^3}{\pi c^3 \left[\exp\left(\frac{\hbar \tilde{\omega}}{k T_{\rm H}}\right)-1\right]},
\end{equation} 
with $n_k/n_j = \exp\left(-\frac{\hbar \tilde{\omega}}{k T_{\rm H}}\right)$. The equilibrium constrains Einstein coefficients as $B_{jk}(\tilde{\omega})=B_{kj}(\tilde{\omega})= \pi c^3 A_{kj}(\tilde{\omega})/(2\hbar \tilde{\omega}^3)$ \cite{1917PhyZ...18..121E}.

\begin{figure}[t]
\centering
 % \begin{centering}
    \includegraphics[width=0.7\textwidth]{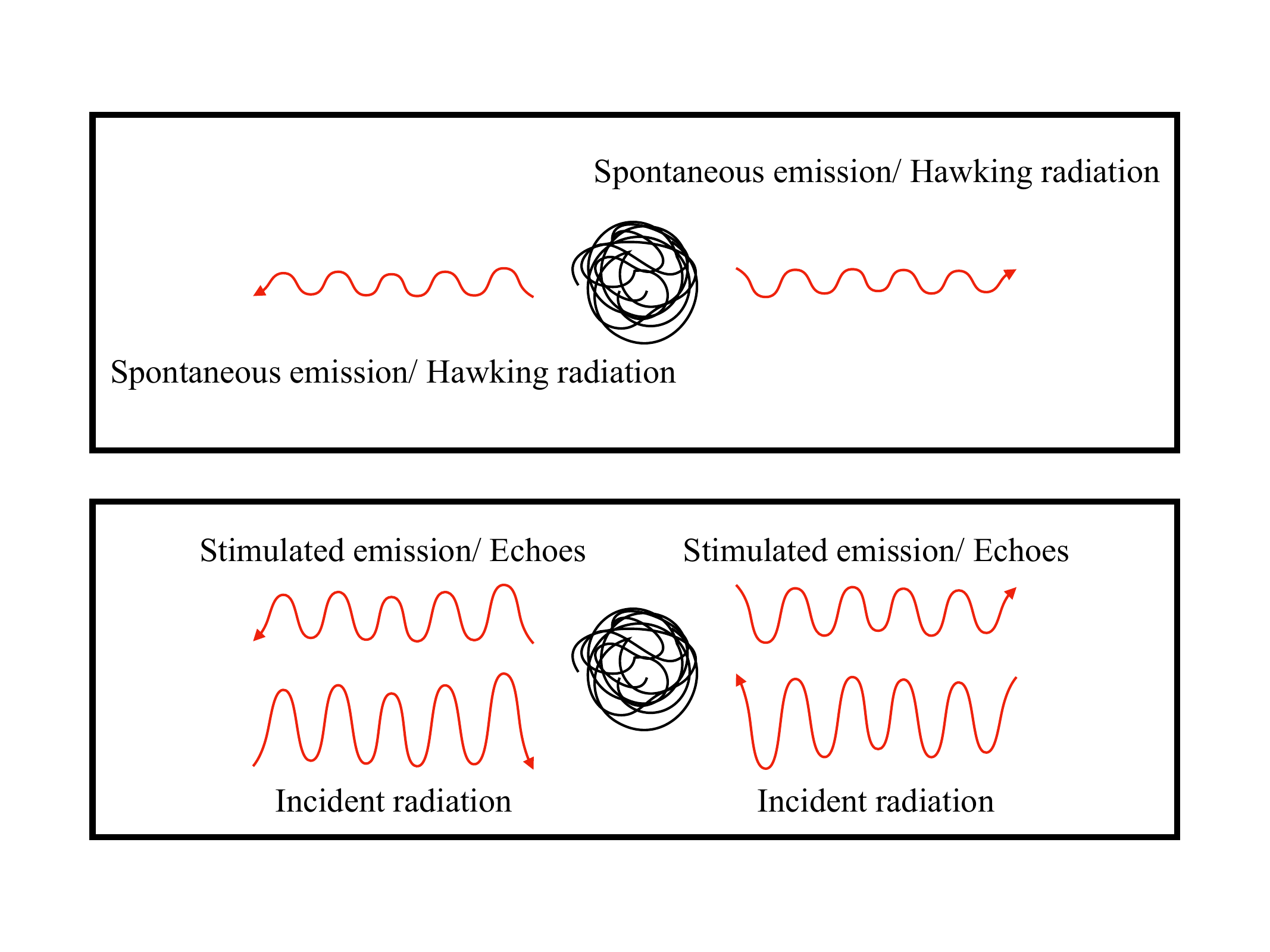}
 % \end{centering}
\caption{Analogy between spontaneous emission and Hawking radiation for isolated BHs, in contrast to stimulated emission caused by incident radiation, that could lead to echoes (see text; note that, for this cartoon we ignore the angular momentum barrier).
}
\label{stimulated}
\end{figure}

Now, imagine a classical incident GW with a much bigger energy flux than Hawking radiation, i.e. $\rho(\tilde{\omega}) \gg \rho_{\rm BB}(\tilde{\omega})$. In this limit, we can ignore spontaneous emission (or Hawking radiation, i.e. 2nd term on the RHS of Eq. \ref{detailed}), to find the reflectivity of the membrane
\begin{equation}
{\mathcal R} = \frac{B_{kj} (\tilde{\omega}) n_k \rho(\tilde{\omega})}{B_{jk} (\tilde{\omega}) n_j \rho (\tilde{\omega})} = \frac{n_k}{n_j} =\exp\left(-\frac{\hbar \tilde{\omega}}{k T_{\rm H}}\right).
\label{reflection1}
\end{equation}
In other words, the slow decay of the quantum BH via Hawking radiation is stimulated by incident GWs for $\hbar \tilde{\omega} \lesssim k T_{\rm H}$, leading to ${\cal O}(1)$ reflectivity (see Fig. \ref{stimulated}). In contrast, in the opposite limit of geometric optics, $\hbar \tilde{\omega} \gg k T_{\rm H}$, the quantum BHs are indeed {\it black}, consistent with the {\it fuzzball complementarity} conjecture \cite{Mathur:2012jk,Chowdhury:2012vd}. 

%%%%%%%%%%%%%%%%%%%%%%%%%
%%%%%%%%%%%%%%%%%%%%%%%%%

For the sake of brevity, in the remainder of this manuscript we shall use natural units with $\hbar=k=c=1$ and we define $r_g \equiv 2 GM$ where $M$ is the mass of the BH.

\subsection{CP-symmetry in the maximally extended spacetime}
We here briefly show that the Boltzmann reflectivity is equivalent to the CP-symmetry of the BH state. Let us consider a mixture of ingoing and outgoing plane waves in Rindler metric (\ref{Rindler}):
\begin{equation}
\psi (x,t) = A_{\text{in}} e^{-i \tilde{\omega} (t + x)} + A_{\text{out}} e^{-i \tilde{\omega} (t-x)}.
\end{equation}
$\psi(x,t)$ can be rewritten in the Minkowski coordinates which are related to the Rindler coordinates as
\begin{equation}
T = \kappa^{-1} e^{\kappa x} \sinh{\kappa t},
X = \kappa^{-1} e^{\kappa x} \cosh{\kappa t},
\end{equation}
where $\kappa$ is the surface gravity of a BH, and then we have
\begin{equation}
\psi (x,t) = A_{\text{in}} [\kappa (T + X)]^{-i\tilde{\omega}/ \kappa}
+ A_{\text{out}} e^{- \pi \tilde{\omega}/ \kappa} [ \kappa (T - X) ]^{i \tilde{\omega}/ \kappa}.
\end{equation}
Imposing the CP-symmetry, $\psi (T,X) = \psi^{\ast} (T, -X)$, one has the following conditions for the coefficients
\begin{align}
A_{\text{in}} &= e^{- \pi \tilde{\omega}/\kappa} A^{\ast}_{\text{out}},\\
A_{\text{out}} &= e^{\pi \tilde{\omega}/\kappa} A^{\ast}_{\text{in}}.
\end{align}
This again leads to the Boltzmann reflectivity of {${\mathcal R} \equiv |A_{\text{out}}/A_{\text{in}}|^2= e^{-\tilde{\omega}/T_{\rm H}}$} {once we properly treat branch-cut}.

Note that this analysis could have been equivalently done in terms of Kruskal/Schwarzschild coordinates. $X \rightarrow -X$ antipodal identification of Kruskal metric is known as an $\mathbbm{RP}^3$ topological geon \cite{Louko:1998dj}. While this spacetime is classically indistinguishable from a BH outside the event horizon, outside quantum measurements can potentially distinguish the two, as the quantum states have different analytic structures \cite{Ng:2017iqh}. However, forming an $\mathbbm{RP}^3$ geon from e.g., stellar collapse requires a non-perturbative change of topology, which can (in principle) happen through quantum tunneling \cite{Mathur:2008kg}.

\subsection{Fluctuation-Dissipation theorem}
{Although the previous two derivations, based on the detailed balance and CP-symmetry, give the Boltzmann reflectivity, they remain ambiguous about the sign of frequency, or the phase of the reflected amplitude. We then next consider a more concrete model motivated by the possible dissipative effects near a BH horizon, which uniquely determines both the amplitude and the phase of reflectivity.}

Classical linear perturbations in BH spacetimes obey the equation  \cite{Regge:1957td,Zerilli:1971wd}
\begin{equation}
\left[\frac{d^2}{dx {}^2} + \tilde{\omega}^2 - V_{\ell}(x) \right] \psi_{\tilde{\omega}} (x) = 0,
\label{RWZ}
\end{equation}
where $V_{\ell}(x)$ is the angular momentum barrier located outside the horizon and $\ell$ is a angular harmonic number. The asymptotic behavior of the (quasinormal) mode function of GWs in the Schwarzschild BH background is
\begin{equation}
\lim_{x \to \pm \infty} \psi_{\tilde{\omega}} = e^{\pm i \tilde{\omega} x}.
\end{equation}

However, we know that quantum effects near BH horizon lead to a thermal behavior at temperature $T_{\rm H}$. According to fluctuation-dissipation theorem \cite{1966RPPh...29..255K}, this should modify the classical field equations via additional fluctuation and dissipation terms, resulting from interaction with quantum/thermal fields. Therefore, we shall posit that Eq. (\ref{RWZ}) is modified to:

% Although the proper wavelength at the distant region is equivalent to $2 \pi/\omega$, in the vicinity of the horizon, the proper wavelength is much shorter than $\ell_{\text{Pl}}$ due to the gravitational blueshift. This is not problematic as the general relativity is a theory for smooth geometries. However, if there exists the microscopic structure on the black hole horizon with the Planck size, plane waves of which the wavelength is about a Planck length would strongly interact with the spacetime microscopic structure. The membrane paradigm \cite{Thorne:1986iy,PhysRevD.18.3598} states that if the structure is disturbed by the blue-shifted incoming waves, the membrane may behave as viscous fluid on the BH. The wave equation for pressure $p$ with viscosity $\nu$ has the form of $(4 \nu/3) \nabla^2 p_t + \nabla^2 p  - \partial_t^2 p = 0$, where the first term gives the viscous effect.
% To simply model the viscous membrane in the wave equation for GWs, we add the dissipation term suppressed by the Planck scale $E_{\text{Pl}}$ in (\ref{RWZ}), which drastically modifies the dispersion relation only at the Planck scale (namely, only near the horizon at which the membrane is located)
\begin{equation}
\left[ -i \frac{\gamma \Omega (x)}{E_{\text{Pl}}} \frac{d^2}{dx {}^2} + \frac{d^2}{dx {}^2} + \tilde{\omega}^2 - V(x) \right]
\psi_{\tilde{\omega}} (x) = \xi_{\tilde{\omega}}(x),
\label{we1}
\end{equation}
where $\xi_{\tilde{\omega}}$ is a stochastic fluctuation field, while $\gamma$ is a dimensionless dissipation parameter, $\Omega (x) \equiv |\tilde{\omega}|/ \sqrt{|g_{00} (x)|}$ is the blueshifted (or proper) frequency, and $E_{\text{Pl}}$ is Planck energy. The form of the dissipation term (which is similar to viscous dissipation for sound waves \cite{1959flme.book.....L}\footnote{The dispersion relation for sound waves that dissipate via fluid (kinematic) viscosity $\nu$ is: $(4 \nu/3) \nabla^2 \partial_tp + c^2_s\nabla^2 p  - \partial_t^2 p = 0$.}) is expected from the fact that gravitational coupling constant is given by $\Omega/E_{\text{Pl}}$. Therefore, dissipation terms coming from gravitational interactions must be suppressed by this factor. In other words, only when the blueshift effect is so intense that the proper frequency is comparable to the Planck energy (i.e. near horizon), $\Omega \sim E_{\text{Pl}}$, the dispersion relation is drastically modified. 

{Moreover, the membrane paradigm \cite{Thorne:1986iy,Jacobson:2011dz}, the fluctuating geometry around a BH \cite{Parentani:2000ts,Barrabes:2000fr} and the minimal length uncertainty principle \cite{Brout:1998ei} lead to the dissipative effects near the apparent horizon. From the point of view of the phenomenology of quantum gravity, constraints on the spacetime viscosity, $\nu$, was also discussed by adding the viscous term of the Navier-Stokes equation, $-i (4/3) \nu \Omega \nabla^2$ \cite{Liberati:2013usa}, as we did in (\ref{we1}). Of course, in lieu of a theory of quantum gravity, there is no clear guiding principle to add dissipation to the dispersion relation, but our choice follows naturally if one follows the analogy between viscous fluids and quantum spacetime.}

According to fluctuation-dissipation theorem, the balance of fluctuation $\xi_{\tilde{\omega}}$ and dissipation should lead to a thermal spectrum for the field $\psi_{\tilde{\omega}}$ \cite{1966RPPh...29..255K}.
Otherwise, for classical fluctuations, $\tilde{\omega}^2 |\psi_{\tilde{\omega}}| \gg |\xi_{\tilde{\omega}}|$, far from the horizon, $\Omega \ll E_{\rm Pl}$, we recover the classical Eq. (\ref{RWZ}). 

Let us calculate the mode function near the horizon where the blueshift effect is most significant.\footnote{We shall focus on large amplitude perturbations, and thus ignore the fluctuation term $\xi_{\tilde{\omega}}$.}
In the near-horizon limit, the exterior metric can be approximated as Rindler
\begin{equation}
ds^2 = e^{2 \kappa x} (-dt^2 + dx {}^2) + dy^2 + dz^2, \label{Rindler}
\end{equation}
where $\kappa =2\pi T_{\rm H}$ is the surface gravity. The modified wave function (\ref{we1}) has an analytic solution near the horizon
\begin{equation}\displaystyle
\lim_{x \to - \infty} \psi_{\tilde{\omega}} (x) = {}_2 F_1 \left[ -i \frac{\tilde{\omega}}{\kappa},i \frac{\tilde{\omega}}{\kappa},1, - i\frac{E_{\text{Pl}} e^{\kappa x}}{\gamma |\tilde{\omega}|}  \right],
\end{equation}
where ${}_2 F_1 (a,b,c,z)$ is the hypergeometric function. In the limit of $x \to - \infty$, this mode function is constant.
%%%%%%%%%%%%%%%%%%%%%%%%%
\begin{figure}[t]
  \begin{center}
    \includegraphics[keepaspectratio=true,height=100mm]{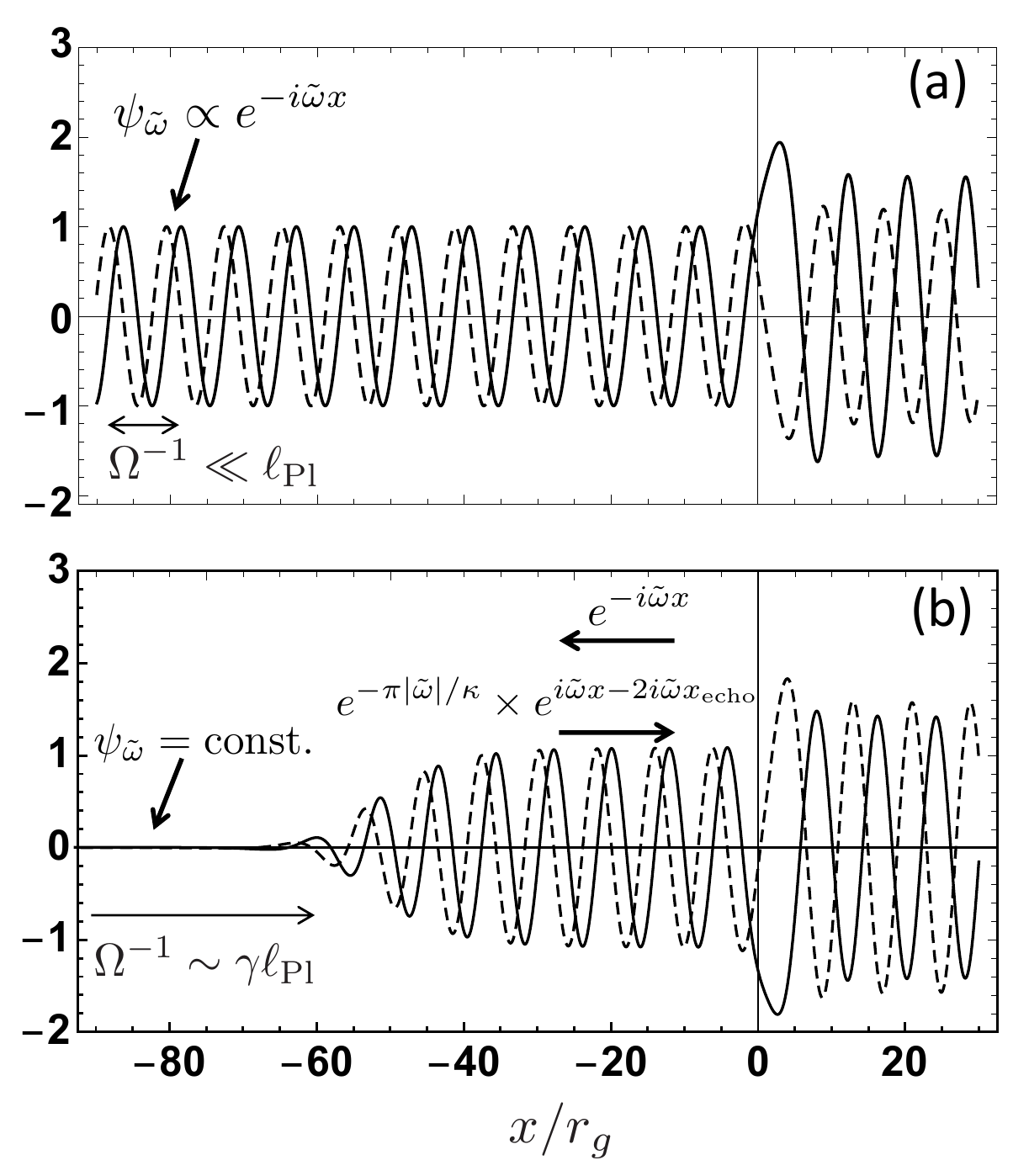}
  \end{center}
\caption{(a) The mode function $\psi_{\tilde{\omega}}(x)$ with the Regge-Wheeler potential, for which the ingoing boundary condition is imposed in the near-horizon limit as in the classical BHs.
(b) The mode function $\psi_{\tilde{\omega}}(x)$ with the Regge-Wheeler potential and the dissipation term added in (\ref{we1}). We take $\gamma = 10$ and $\kappa = (2r_g)^{-1}$. In both (a) and (b) we take $\ell = 2$, $s = -2$, and
$r_g \tilde{\omega} = 0.8$. 
Solid and dashed lines show $\Re[\psi_{\tilde{\omega}}]$ and $\Im[\psi_{\tilde{\omega}}]$, respectively.
}
\label{mode}
\end{figure}
%%%%%%%%%%%%%%%%%%%%%%%%%
The physical meaning of this boundary condition is that the energy flux carried by the ingoing GWs cannot penetrate the horizon, and is either absorbed or reflected. In this sense, this boundary condition is consistent with the picture according to a distant observer in the context of BH complementarity \cite{Susskind:1993if} or the membrane paradigm \cite{Thorne:1986iy,PhysRevD.18.3598}, in which there is virtually no BH interior to propagate into.
In the range of $- \log \left[E_{\text{Pl}} / (\gamma |\tilde{\omega}|) \right] \ll \kappa x \ll -1$, the function can be expressed by the superposition of outgoing and ingoing modes
\begin{align}
\psi_{\tilde{\omega}} &= e^{\pi |\tilde{\omega}|/(2 \kappa)}A e^{-i \tilde{\omega} x} + e^{-\pi |\tilde{\omega}|/(2 \kappa)} A^{\ast} e^{i \tilde{\omega} x},\\
A & \equiv \frac{\left( \frac{\gamma |\tilde{\omega}|}{E_{\text{Pl}}} \right)^{i \tilde{\omega}/ \kappa} \Gamma (-2i \tilde{\omega}/ \kappa)}{\Gamma (-i \tilde{\omega}/\kappa) \Gamma (1-i \tilde{\omega}/\kappa)}.
\end{align}
Therefore, we again recover the Boltzmann reflectivity (\ref{reflection1}):
\begin{equation}
{\mathcal R} = \left| \frac{e^{-\pi |\tilde{\omega}|/(2 \kappa)}A^{\ast}}{e^{\pi |\tilde{\omega}|/(2 \kappa)}A}\right|^2=e^{- |\tilde{\omega}|/T_{\rm H}}. \label{ref_2}
\end{equation}
For the detailed derivation in the Kerr background, see Appendix \ref{app:boltzmann}.

Remarkably, the flux reflectivity ${\cal R}$ is independent of the dissipation parameter $\gamma$ in Eq. (\ref{we1}), even though the approximate position of the reflection $x_{\rm echo}$ (where $\gamma \Omega \sim E_{\rm Pl}$), and hence echo time delays  \cite{Oshita:2018fqu}, does depend on it:\footnote{{Note that since we can only derive the phase of reflectivity from our Equation (\ref{we1}), its derivation is more model-dependent than that of Boltzmann reflectivity for the energy.}}
\begin{equation}
\Delta t_{\rm echo} = 2|x_{\rm echo}|= 2\kappa^{-1}\ln\left[E_{\rm Pl}/(\gamma |\tilde{\omega}|)\right].
\label{echo_time}
\end{equation}

\subsection{Relation between Reflectivity and Viscosity}
Motivated by the fluid/gravity duality and the membrane paradigm \cite{Thorne:1986iy,PhysRevD.18.3598}, we can relate BH horizon reflectivity to the membrane fluid viscosity.  {In the Appendix \ref{app:viscosity}, we show that the ratio of outgoing to ingoing GW amplitude in the near-horizon limit is related to fluid viscosity: }
\begin{equation}\label{Aoutin_eta}
    \frac{A_{\rm out}}{A_{\rm in}} =\frac{1-16 \pi G \eta}{1+ 16 \pi G \eta} e^{-2i\tilde{\omega} x}.
\end{equation}
Hence, the reflectivity is related to viscosity via ${\cal R}=\left(\frac{1-16 \pi G \eta}{1+16 \pi G \eta}\right)^2$. For Boltzmann reflectivity (\ref{reflection1}), this yields
\begin{equation}
    \eta=\frac{1}{16 \pi G} \tanh\left(\frac{\hbar |\tilde{\omega}|}{4k  T_{\rm H}}\right).
\end{equation} 
Therefore, we recover the standard membrane paradigm shear viscosity $\frac{1}{16 \pi G}$ at high frequencies, but $\eta$ approaches zero linearly at low frequencies. %This implies that the reflectivity is more dominant than viscousity for a lower frequency.

\section{GW Echoes and absence of ergoregion instability}
\label{sec:ECHO}
A non-vansihing horizon reflectivity will lead to echoes  from the ringdown of a perturbed BH (e.g., \cite{Cardoso:2016rao,Cardoso:2016oxy,Nakano:2017fvh,Wang:2018gin,Burgess:2018pmm,Oshita:2018fqu}). Indeed, tentative (albeit controversial) evidence for these echoes have been claimed in the literature \cite{Abedi:2016hgu,Conklin:2017lwb,Abedi:2018npz}. Here we outline the basic features of GW echoes from Boltzmann reflectivity, while a companion paper examines these predictions and implications for BH quasinormal modes in more detail \cite{echo_QBH}. We further show that (consistent with current observational bounds \cite{Barausse:2018vdb}), ergoregion instability is not expected for spinning BHs, due to imperfect reflectivity.

% Nakano \textit{et al.} \cite{Nakano:2017fvh} discussed howling echo-GWs from a spinning BH with a perfect-reflection boundary since the superradiant instability would appear in such a case. However, the reflection rate of BHs we have discussed so far has the frequency dependence and is not perfect reflection for $\omega > 0$. In the following, we show that the superradiant instability does not appear in the rotating quantum BHs.

In order to investigate how the dissipation term change the ringdown GWs propagating from a spinning quantum BH, we start with the Sasaki-Nakamura (SN) equation\footnote{One can choose other wave equations, e.g. Chandrasekhar-Detweiler equation \cite{Chandrasekhar:1976zz}, which also has its short range angular momentum barrier. In either case, the qualitative result does not change since the detail of angular momentum barrier is irrelevant for the derivation of (\ref{ref_2}).} \cite{Sasaki:1981sx} including the dissipation term (hereinafter referred to as the modified SN equation).
{Although there is no unique choice of wave equation around a Kerr BH, we here choose  the SN equation that describes the wave propagation in the co-rotating coordinates, and so there is no time-asymmetry due to the rotation of background spacetime. As such, the only terms that break time-asymmetry are dissipative terms near horizon, where propagation is governed by near-horizon Rindler geometry, and is fully fixed by surface gravity $\kappa=2\pi T_{\rm H}$.}

In the near-horizon limit, one can obtain the mode function (as we did before) by satisfying the no-flux condition ($\psi_{\tilde{\omega}} (x) = \text{const.}$ for $x \to - \infty$), yielding the Boltzmann reflectivity of Eq. (\ref{ref_2}) ({see the Appendix \ref{app:boltzmann}}). 
%%%%%%%%%%%%%%%%%%%%%%%%%
\begin{figure}[t]
  \begin{center}
    \includegraphics[keepaspectratio=true,height=75mm]{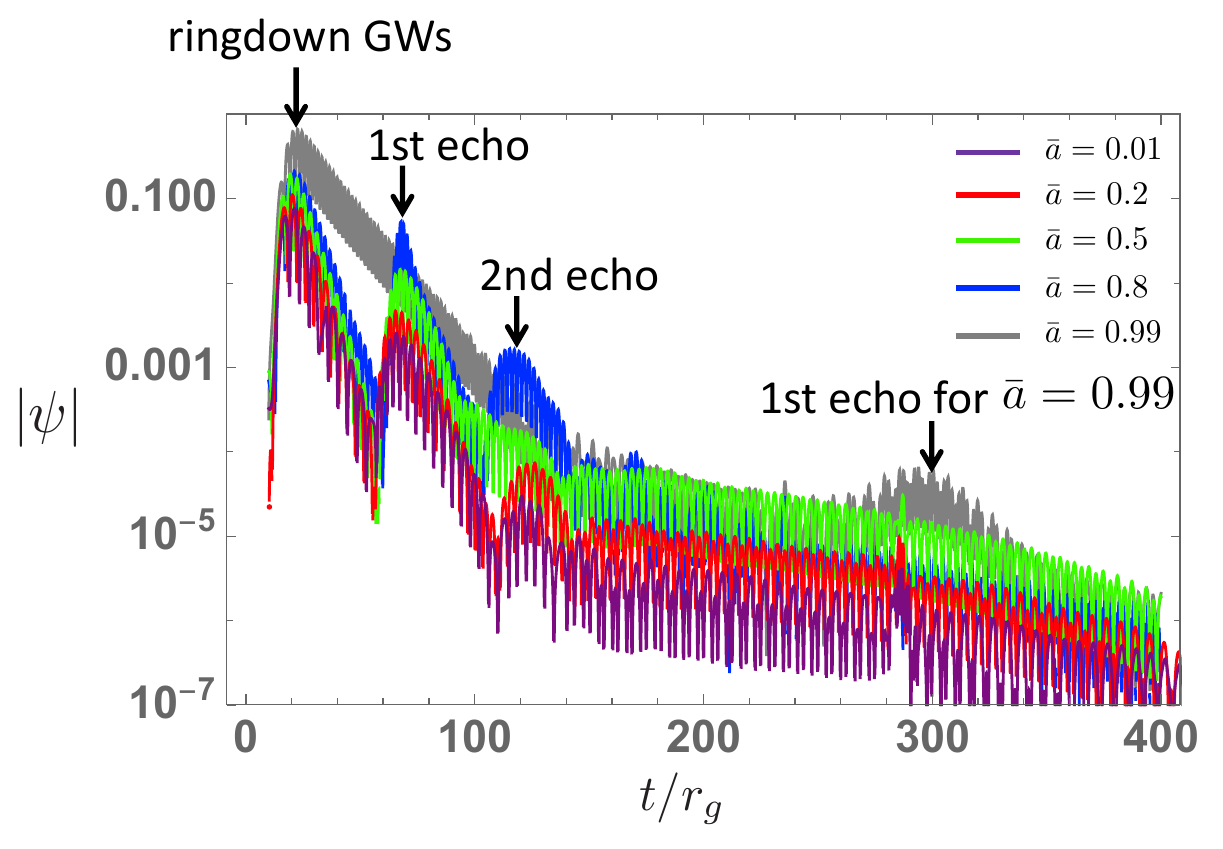}
  \end{center}
\caption{
The ringdown signals followed by the echo GWs with $\sigma = 2 r_g$, $\sigma_{\text{h}} = 0.45 r_g$, and $x_c = 0$ seen by an observer at $x = 25 r_g$. For illustrative purposes, we take $r_g = 10^5 \ell_{\text{Pl}}$ and $\gamma = 10$. The spin parameter is $\bar{a}=0.01$ (purple), $\bar{a}=0.2$ (red), $\bar{a}=0.5$ (green), $\bar{a}=0.8$ (blue), and $\bar{a} = 0.99$ (gray).
}
\label{echo}
\end{figure}
%%%%%%%%%%%%%%%%%%%%%%%%%
%%%%%%%%%%%%%%%%%%%%%%%%%
\begin{figure}[b]
  \begin{center}
    \includegraphics[keepaspectratio=true,height=65mm]{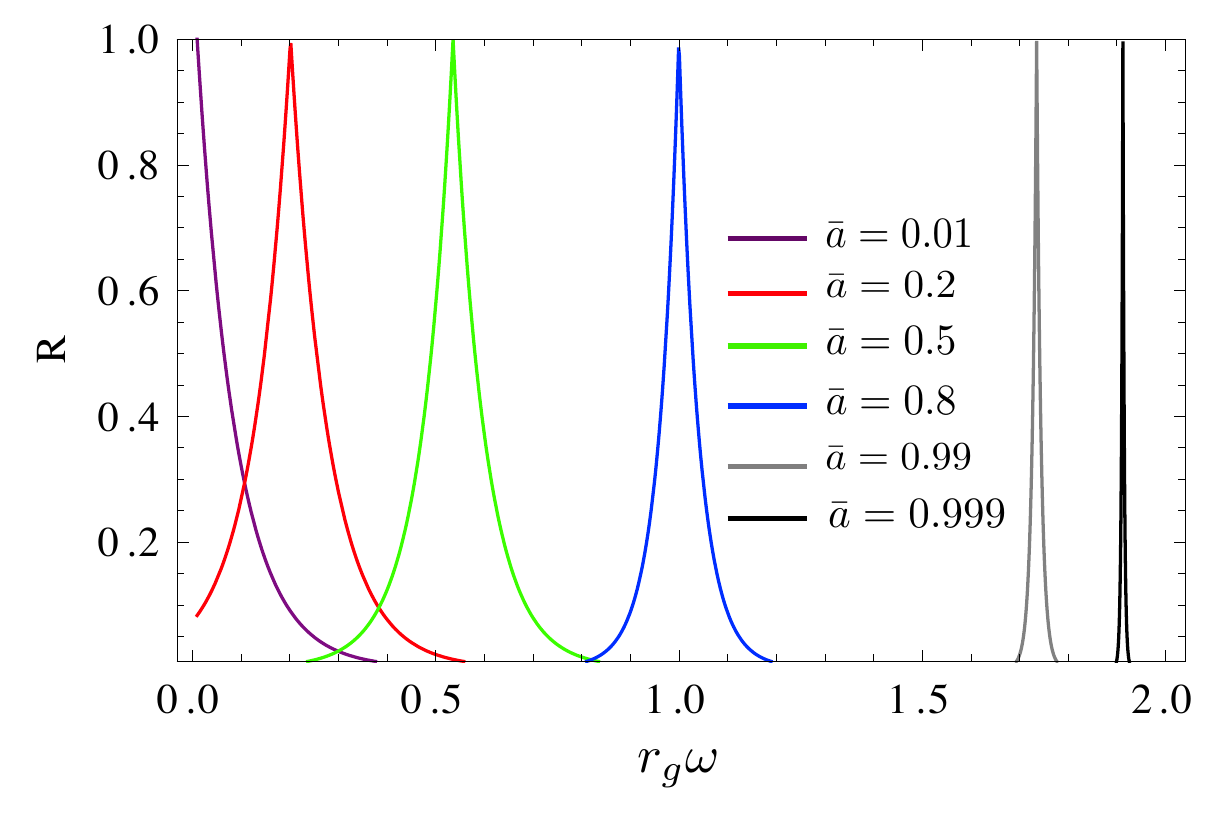}
  \end{center}
\caption{The frequency dependence of the Boltzmann reflectivity (for $m=2$ mode) assuming a spinning BH with spin parameters $\bar{a}=0.01$, $0.2$, $0.5$, $0.8$, $0.99$, and $0.999$.}
\label{reflection}
\end{figure}

An example of a full waveform can be obtained by starting with a Gaussian wavepacket 
\begin{equation}
\psi(x,0) = \exp\left[-\frac{(x-x_c)^2}{\sigma^2}-\frac{ix}{\sigma_{\rm h}}\right],
\dot{\psi}(x,0) = 0, 
\end{equation}
where a dot denotes the derivative with respect to $t$, $\sigma$ and $\sigma^{-1}_{\rm h}$ characterize the width and mean wavenumber of the wavepacket, while $x_c$ is its initial position.
Fig. \ref{echo} shows numerical integration of the modified SN equation to find the GW strain amplitude seen by a distant observer. In addition to the original ringdown, as expected, we see echoes with a time-delay given by Eq. (\ref{echo_time}), with a spin-dependent amplitude. In particular, in the extremal limit $\bar{a}\to 1$ for fixed initial conditions, {where $\bar{a}$ is the non-dimensional spin parameter}, echoes are highly suppressed (and delayed), since $T_{\rm H} \to 0$, and thus the reflectivity is exponentially suppressed except for a narrow range around {$\omega = m\Omega_{\rm H} \equiv \frac{m \bar{a}}{(1+\sqrt{1-\bar{a}^2})r_g}$ (see Fig. \ref{reflection}), where $\omega \equiv \tilde{\omega} + m\Omega_{\rm H}$ and $m$ is the azimuthal number.} This is enough to suppress ergoregion instability, even for a rapidly spinning BH \cite{Maggio:2018ivz}.

\section{Conclusions}
We have provided three independent derivations for a Boltzmann reflectivity of quantum BH horizons, $\mathcal{R} = e^{- |\tilde{\omega}|/T_{\rm H}}$, based on
\begin{enumerate}
    \item Thermodynamic detailed balance,
     \item CP-symmetry, or $\mathbbm{RP}^3$ topology, of extended BH spacetime, and
    \item Fluctuation-dissipation theorem.
\end{enumerate}
Therefore, although a concrete picture of microscopic structure of a quantum BH is still missing, macroscopic properties such as entropy, temperature, and now, its energy flux reflectivity, may be independent of the details. 

Assuming this universal property of the quantum BHs,
we numerically investigated the GW echoes and showed that the echo is strongly suppressed, and delayed, for a rapidly spinning BH, $0 < 1-\bar{a} \ll 1$, due to the decrease of its Hawking temperature. This leads to the absence of the ergoregion instability since the frequency dependence of the Boltzmann reflectivity is sharply peaked around $\omega \simeq m\Omega_{\rm H} \pm T_{\rm H}$ and is exponentially suppressed outside this range. Finally, we discussed the implication for the fluid viscosity in the membrane paradigm, finding that it should vanish at low frequencies.

The synergy of our three, seemingly independent, derivations all leading to a Boltzmann reflectivity may help us draw a clearer picture of what a quantum black hole might look like (as long as we assume validity of linear perturbation theory): 

The central assumption underlying our key result, Eq. (\ref{reflection1}), is that black holes are not classical spacetimes, but rather quantum objects, obeying standard rules of unitary quantum mechanics and thermodynamics. For one, this implies that they cannot be perfectly absorbing, as it would violate unitarity. Furthermore, as typical de-excitation of BH state leads to emission of Hawking photons/gravitons, it is reasonable to assume that typical absorption happens at similar frequencies, and thus photons/gravitons at much lower frequencies, $\hbar\tilde{\omega} \ll k T_{\rm H}$ cannot excite {\it local} membrane degrees of freedom. As such, they should be reflected, which is exactly what is predicted by Boltzmann reflectivity (\ref{reflection1}). In other words, quantum BHs must be ``optically thin'' at $\hbar\tilde{\omega} \ll k T_{\rm H}$.

The latter point is further underscored by the fluctuation-dissipation derivation, as lower frequency waves are reflected farther away from the horizon, where gravity is weaker. It also suggests that this universality might be a low-energy property, as higher powers of $\Omega/E_{\rm Pl}$ can modify the dispersion relation at higher frequencies \cite{Oshita:2018fqu}. However, this may not be a significant correction since reflectivity is already highly suppressed for $\hbar\tilde{\omega} \gtrsim k T_{\rm H}$.

Finally, the equivalence of the CP-symmetry, or $\mathbbm{RP}^3$ topology, of the BH spacetime with Boltzmann reflectivity suggests that a drastic transition from classical to quantum BHs is necessary. This is clearly not something that would emerge from e.g. classical collapse of a star, and requires a change of topology through non-perturbative quantum tunneling, similar to what is advocated in the fuzzball proposal \cite{Mathur:2008kg}. Therefore, even though we rely on perturbation theory outside stretched horizon in our analysis, from a global perspective, a non-perturbative transition is necessary to take a classical BH to a quantum BH spacetime. 

% However, we must stress that this is {\it not} a mathematical proof, as it is based on linear perturbation theory, and thus can be violated, depending on a more complete model for quantum black holes.   

Let us end by noting that, as we discuss in a companion paper \cite{echo_QBH},  our concrete predictions for GW echoes from quantum BHs are imminently testable using current and upcoming GW observations. May we suggest that we hold other proposals for quantum black holes to the same standard?!

\appendix

\section{Derivation of $e^{-|\tilde{\omega}|/T}$ from the modified dispersion relation}
\label{app:boltzmann}
Here we show the details of the derivation of $e^{-|\tilde{\omega}|/T}$ in the Kerr spacetime by starting with the analytic solution of the modified SN equation. The absolute value of the frequency is very important since the ergoregion instability can be induced if the reflectivity was given by $e^{- \tilde{\omega}/T}$. 
Let us start with the modified SN equation:
\begin{equation}
\left( \frac{- i \gamma |\tilde{\omega}|}{\tilde{F} \sqrt{\delta (r)} E_{\text{Pl}}} \frac{d^2}{dx {}^2} + \frac{d^2}{dx {}^2} - {\mathcal F} \frac{d}{dx} - {\mathcal U} \right) \psi_{\tilde{\omega}} =0,\label{SNeq}
\end{equation}
in tortoise coordinate:
\begin{align}
\begin{split}
x &\equiv \int dr \frac{r^2+a^2}{r^2\delta(r)} \\
&= r+ \frac{r_g r_+}{r_+-r_-} \ln{\frac{r-r_+}{r_g}} - \frac{r_g r_-}{r_+-r_-} \ln{\frac{r-r_-}{r_g}},
\end{split}
\end{align}
where $a \equiv GM \bar{a}$ is the spin parameter, $r_{\pm} \equiv GM \pm \sqrt{(GM)^2 -a^2}$, the forms of ${\mathcal F}$ and ${\mathcal U}$ can be found in \cite{Sasaki:1981sx}, {and $\tilde{F} \sqrt{\delta(r)}$ is the blue shift factor of Kerr spacetime derived in \cite{NouriZonoz:1998td}}
\begin{align}
\begin{split}
&\tilde{F} \equiv \sqrt{\frac{r^2 (r^2+ a^2 \cos^2{\theta})}{(r^2+a^2) (r^2+a^2 \cos^2{\theta})+a^2  r_g r \sin^2{\theta}}},\\
& \times \frac{a^2 (r^2+a^2) \cos^2{\theta} +r^2 (r^2 + a^2 + r_g a^2 \sin^2{\theta}/r)}{(r^2+a^2) (r^2+a^2 \cos^2{\theta})},
\end{split}\\
&\delta^{1/2} \equiv \sqrt{1-r_g/r +a^2/r^2},
\end{align}
%{Since freely falling GWs near the horizon co-rotates with the Kerr geometry, one should take the blue-shift factor in terms of the co-rotating frame although the factor in terms of Boyer-Lindquist coordinates, which corresponds to the rest frame, gives a different blue-shift factor.}
For spinning BHs, with the Hawking temperature is
\begin{equation} 
T \equiv \frac{\kappa_+}{2 \pi} = \frac{1}{2\pi r_g} \left(\frac{\sqrt{1-\bar{a}^2}}{1+\sqrt{1-\bar{a}^2}}\right),
\end{equation}
where $\kappa_+$ is the surface gravity at the outer horizon $r=r_+$, while the horizon-frame frequency $\tilde{\omega}$ is related to the frequency seen by the distant observer $\omega$ via
\begin{equation}
    \tilde{\omega} = \omega - m \Omega_{\rm H},~~
\Omega_{\rm H} =  \frac{\bar{a}}{(1+\sqrt{1-\bar{a}^2})r_g}.
\end{equation}
Here, $\Omega_{\rm H}$ is the angular velocity of the horizon, and $m$ is the azimuthal angular momentum number (=2 for dominant mode of BH ringdown perturbations).

In the near horizon limit ($x \to - \infty$), the blue-shift factor reduces to
\begin{equation}\displaystyle
\lim_{x \to - \infty} \tilde{F} (r, \theta) \sqrt{\delta} = \tilde{F} (r= r_+, \theta) C e^{\kappa_+ x},
\end{equation}
where $C$ has the form of
\begin{equation}
C \equiv  \exp \left[ \frac{1}{2} \frac{\sqrt{1-\bar{a}^2}}{r_+^2/r_g^2 +\bar{a}^2/4} \left( - r_+/r_g + \frac{r_-^2/r_g^2 +\bar{a}^2/4}{2\sqrt{1-\bar{a}^2}} \log{(1-\bar{a}^2)} \right) + \frac{1}{2} \log \sqrt{1-\bar{a}^2} -\log(r_+/r_g) \right],
\end{equation}
and the SN equation reduces to the following form:
\begin{equation}
\left(-i \frac{\gamma |\tilde{\omega}|}{C \tilde{F} E_{\text{Pl}}} e^{-\kappa_+ x} \frac{d^2}{dx^2} + \frac{d^2}{dx^2} - \tilde{\omega}^2 \right) \psi_{\tilde{\omega}} = 0.
\label{SN_near_horizon}
\end{equation}
The solution of (\ref{SN_near_horizon}) which satisfies the aforementioned boundary condition is
\begin{equation}\displaystyle
\lim_{x \to -\infty} \psi_{\tilde{\omega}} = {}_2 F_1 \left[ -i \frac{\tilde{\omega}}{\kappa_+}, i \frac{\tilde{\omega}}{\kappa_+}, 1, -i \frac{C \tilde{F} E_{\text{Pl}} e^{\kappa_+ x}}{\gamma |\tilde{\omega}|} \right],
\end{equation}
and one can read that in the intermediate region, $- \kappa_+^{-1} \log \left[C \tilde{F} E_{\text{Pl}} / (\gamma |\tilde{\omega}|) \right] \ll x \ll -r_g$, $\psi_{\tilde{\omega}}$ can be expressed as the superposition of outgoing and ingoing modes
\begin{align}
\psi_{\tilde{\omega}}=
\begin{cases}
e^{\pi \tilde{\omega}/(2 \kappa_+)}A_+ e^{-i \tilde{\omega} x} + e^{-\pi \tilde{\omega}/(2 \kappa_+)} A_+^{\ast} e^{i \tilde{\omega} x} \ &\text{for} \ \tilde{\omega} > 0,\\
e^{-\pi \tilde{\omega}/(2 \kappa_+)}A_- e^{-i \tilde{\omega} x} + e^{\pi \tilde{\omega}/(2 \kappa_+)} A_-^{\ast} e^{i \tilde{\omega} x} \ &\text{for} \ \tilde{\omega} < 0,
\end{cases}
\end{align}
where $A_{\pm}$ has the form of
\begin{align}
&A_{\pm} \equiv \frac{\Gamma (-2i \tilde{\omega}/ \kappa_+)}{\Gamma (-i \tilde{\omega}/\kappa_+) \Gamma (1-i \tilde{\omega}/\kappa_+)} e^{i \tilde{\omega} x_{\text{echo}}},\\
&\text{with} \ \ x_{\text{echo}} = \frac{1}{\kappa_+} \log{\left[ \frac{\gamma |\tilde{\omega}|}{C \tilde{F} (\theta) E_{\text{Pl}}} \right]}.\label{x_echo_kerr}
\end{align}
Therefore, the energy reflectivity is given by
\begin{equation}
R =
\begin{cases}
\left| \frac{e^{- \pi \tilde{\omega}/(2 \kappa_+)} A_+^{\ast}}{e^{\pi \tilde{\omega}/(2 \kappa_+)} A_+} \right|^2 = e^{- 2\pi \tilde{\omega}/\kappa_+} \ \ \text{for} \ \ \tilde{\omega} > 0,\\
\left| \frac{e^{ \pi \tilde{\omega}/(2 \kappa_+)} A_-^{\ast}}{e^{-\pi \tilde{\omega}/(2 \kappa_+)} A_-} \right|^2 = e^{ 2\pi \tilde{\omega}/\kappa_+} \ \ \text{for} \ \ \tilde{\omega} < 0,
\end{cases}
\end{equation}
{where we used $|A_{\pm}^{\ast}/A_{\pm}| = 1$,} and finally we obtain $R = e^{-|\tilde{\omega}|/T}$.

\begin{figure}[h]
  \begin{center}
    \includegraphics[keepaspectratio=true,height=70mm]{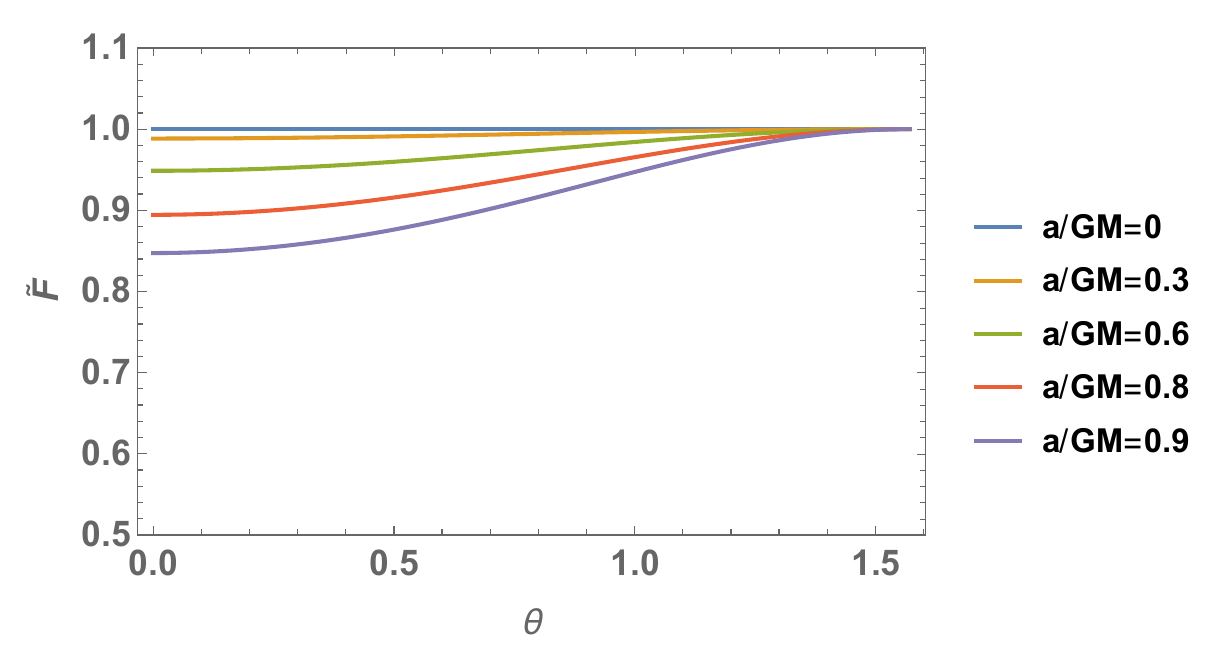}
  \end{center}
\caption{Plot of $\tilde{F} (r=r_+, \theta)$ for various values of spin.
}
\label{pic2}
\end{figure}

Let us note that, in deriving Equation (\ref{SN_near_horizon}) from (\ref{SNeq}) in the near-horizon limit, we have ignored the angular momentum barrier terms ${\mathcal F} \frac{d}{dx} - {\mathcal U}$, which are exponentially suppressed near horizon and are negligible as long as $|x_{\rm echo}| \sim \kappa_+^{-1}\ln\left[E_{\rm Pl}/(\gamma |\tilde{\omega}|)\right] \ll \kappa^{-1}_+$. Given that Astrophysical gravitational wave frequencies are $\sim 10^2$ Hz, while Planck frequency/energy is $10^{43}$ Hz, this means that our derivation of Boltzmann reflectivity is independent of the exact value of $\gamma$, as long as $\gamma \ll 10^{41}$. 
{Furthermore, the trans-Planckian frequency is involved at $x=x_{\text{echo}}$ when $\gamma \lesssim 1$, in which the semi-classical treatment might break down and so the range of $\gamma$ would be restricted to $\gamma \gtrsim 1$ to avoid this. Therefore, our calculation may only be valid for
\begin{equation}
1 \lesssim \gamma \ll 10^{41}.
\end{equation}
We should also remark on the $\theta$-dependence of the blue-shift factor $\tilde{F}$. The $\theta$-dependence implies that the separability of the Teukolsky equation with our modification term breaks down. However, as is shown in FIG. \ref{pic2}, $\tilde{F}$ only has a small angular variation and appears inside the $\log$ in $x_{\text{echo}}$ (see Eq. \ref{x_echo_kerr}). The effect on $\Delta t_{\rm echo}$ is $<0.1\%$, and thus can be safely ignored.} {We note, however, that the full problem of the separability of the modified wave equation may be more complicated and our argument is only approximately valid in the specific case.}
%%%%%%%%%%%%%%%%%%%%%%%%%
%%%%%%%%%%%%%%%%%%%%%%%%%

\section{Relation between Reflectivity and Viscosity in the Membrane Paradigm}
\label{app:viscosity}
In this section, we provide the derivation of Equation which relates the reflectivity of the ``horizon'' to viscosity in the context of membrane paradigm. For simplicity, we apply Regge-Wheeler formalism with axial axisymmetric perturbation, $\delta g_{\mu \nu}$, in the Schwarzschild spacetime $g_{\mu\nu} = g^{\rm Sch}_{\mu\nu}(r) +\delta g_{\mu\nu}(r,\theta,t)$, where
\begin{align}
\delta g_{t \phi}&= \epsilon e^{- i \tilde{\omega} t} h_0(r) y(\theta),\label{htphi} \\
\delta g_{r \phi}&= \epsilon e^{- i \tilde{\omega} t} h_1(r) y(\theta),
\end{align}
while other $\delta g_{\mu \nu}$ components vanish, and $\epsilon \ll 1$ controls the order of perturbation. Following \cite{Jacobson:2011dz}, we can define the Brown-York stress tensor via the Israel junction conditions $K_{ab}-K h_{ab} = - 8 \pi G T_{ab}$ for the membrane now standing at $r=r_0+ \epsilon R[t,\theta]$, where $h_{ab}$ is the induced metric on the membrane, $K_{ab}$ is its extrinsic curvature, and $r_0$ is its unperturbed position. Here, the indexes $\mu, \nu$ run over $(t,r,\theta,\phi)$, while $a, b$ run over $(t,\theta,\phi)$. $T_{ab}$ is assumed to be the energy momentum tensor of a viscous fluid:
\begin{align}
     T_{ab}= [\rho_0+ \epsilon \rho_1(t,\theta) ] u_{a} u_{b}+ \nonumber\\ [ p_0+ \epsilon p_1(t,\theta)-\zeta \Theta] \gamma_{ab} -2 \eta \sigma_{ab},\\
     \sigma_{ab}=\frac{1}{2}(u_{a;c} \gamma^c_b+u_{b;c}\gamma^c_a- \Theta \gamma_{ab}), \\
     \gamma_{ab} \equiv h_{ab}+u_{a}u_{b}, \quad \Theta \equiv u^a_{;a},\label{theta}
\end{align}
where $\rho_0$ and $p_0$ ($\rho_1$ and $p_1$) are background (perturbation in) membrane density and pressure, while $u_a$, $\eta$ and $\zeta$ are fluid velocity, shear viscosity, and bulk viscosity, respectively. 

Plugging Eqs. (\ref{htphi}-\ref{theta}) into the the Israel junction condition and expanding to 1st order in $\epsilon$, we find
% \begin{align}
%      \rho_0[r_0]&= -\frac{\sqrt{f[r_0]}}{4 \pi r_0 },\\
%      p_0[r_0]&= \frac{\sqrt{f[r_0]} (g[r_0]+r_0 g'[r_0])}{8 \pi r_0 g[r_0]},
% \end{align}
%  , where $g[r_0] = (1-r_0/2)^{1/2}$ and $f[r_0] = 1-r_0/2$. Next order solution from $T_{\theta \phi}$ assuming $u_{\phi}=0$ gives 
 \begin{equation}
     \tilde{\omega} h_1(r) =-8 i \pi G \eta [h_1(r) +(r-r_g)h'_1(r)]. 
     \label{boundary}
 \end{equation}
We can further use $\psi_{\tilde{\omega}} = \frac{1}{r} \left(1-\frac{r_g}{r}\right) h_1(r)$ in the tortoise coordinate $x=r+r_g \log[r/r_g -1]$ to rewrite Eq. (\ref{boundary}) as
\begin{equation}
    \tilde{\omega}\psi_{\tilde{\omega}} = 16 i \pi G \eta \frac{\partial \psi_{\tilde{\omega}}}{ \partial x}.
    \label{boundaryrw}
\end{equation}
For the standard horizon with a purely ingoing boundary condition $\psi_{\tilde{\omega}} \propto e^{-i\tilde{\omega} x}$, plugging into Eq. (\ref{boundaryrw}), $ \eta =\frac{1}{16 \pi G} $ is a constant, which is what we get from the standard membrane paradigm. If instead we have a partially reflective membrane $\psi_{\tilde{\omega}} = A_{out} e^{i\tilde{\omega} x} + A_{in} e^{-i\tilde{\omega} x} $, we find 
\begin{equation}
    \frac{A_{\rm out}}{A_{\rm in}} =\frac{1-16 \pi G \eta}{1+ 16 \pi G \eta} e^{-2i\tilde{\omega} x}.
\end{equation}

\acknowledgments

We thank Jahed Abedi, Vitor Cardoso, Randy S. Conklin, Steve Giddings, Bob Holdom, Elisa Maggio, Samir Mathur, Emil Motolla, Jing Ren, Sergey Sibiryakov, Rafael Sorkin,  Daichi Tsuna, Huan Yang, Yuki Yokokura, and Aaron Zimmerman for helpful comments and discussions. We also thank all the participants in our weekly group meetings for their patience during our discussions. This work was supported by the University of Waterloo, Natural Sciences and Engineering Research Council of Canada (NSERC), the Perimeter Institute for Theoretical Physics, and the JSPS Overseas Research Fellowships (N. O.). Research at the Perimeter Institute is supported by the Government of Canada through Industry Canada, and by the Province of Ontario through the Ministry of Research and Innovation.
Most of this work was done while N. O. was at the Research Center for the Early Universe (RESCEU) in the University of Tokyo and N. O. thanks to all members in RESCEU.


\end{document}